# Deciphering the Intense Post-Gap Absorptions of Monolayer Transition Metal Dichalcogenides

Jinhua Hong[†,*], Masanori Koshino[†], Ryosuke Senga[†], Thomas Pichler[‡], Hua Xu[§], Kazu Suenaga[†,∥,*]

[†]Nanomaterials Research Institute, National Institute of Advanced Industrial Science and Technology (AIST), Tsukuba 305-8565, Japan

[‡]Faculty of Physics, University of Vienna, Strudlhofgasse 4, A-1090 Vienna, Austria

[§]Key Laboratory of Applied Surface and Colloid Chemistry, School of Materials Science and Engineering, Shaanxi Normal University, Xi'an 710119, P. R. China

[∥]The Institute of Scientific and Industrial Research (ISIR-SANKEN), Osaka University, Ibaraki 567-0047, Japan

[*]E-mail: jinhuahong436@gmail.com (J. H.), or suenaga-kazu@sanken.osaka-u.ac.jp (K. S.)

## ABSTRACT

**Rich valleytronics and diverse defect-induced or interlayer pre-bandgap excitonics have been extensively studied in transition metal dichalcogenides (TMDCs), a system with fascinating optical physics. However, more intense high-energy absorption peaks (~ 3 eV) above the bandgaps used to be long ignored and their underlying physical origin remains to be unveiled. Here, we employ momentum resolved electron energy loss spectroscopy to measure the dispersive behaviors of the valley excitons and intense higher-energy peaks at finite momenta. Combined with accurate Bethe Salpeter equation calculations, non-band-nesting transitions at Q valley and at midpoint of KM are found to be responsible for the high-energy broad absorption peaks in tungsten dichalcogenides and present spin polarizations similar to A excitons, in contrast with the band-nesting mechanism in molybdenum dichalcogenides. Our experiment-theory joint research will offer insights into the physical origins and manipulation of the intense high-energy excitons in TMDCs-based optoelectronic devices.**

Atomically thin semiconducting transition metal dichalcogenides (TMDCs) have long been regarded as an ideal 2D platform exhibiting rich structural, opto/electronic[1] and valleytronic[2-4] properties. In the low-energy end of their optical absorption, lots of efforts have been devoted for the discovery of spin-valley coupled valleytronics,[2-4] dark excitons,[5,6] defect induced single photon emission,[7-9] and interlayer Moiré excitons[10,11] in hetero-bilayers. All these optical phenomena occur either at or below the energy of A exciton (typically 1 ~ 2 eV). Contrastingly, the absorptions at higher energies show much higher intensity than the A excitons (Figure 1a) but receive much less attention and remains almost untouched or evaded by researchers. Moreover, all reported optical measurements detect electronic excitations with zero momentum transfer ($q$) only. Non-zero-$q$ excitation is also of great importance to probe exciton dispersions and will give direct hints to track $k$-space origins of the unknown high-energy electronic transitions, as illustrated in Figure 1b. Therefore, direct experimental exploration of high-energy excitons is highly demanding, as they play a more prominent role in the optical absorption and charge carrier dynamics in TMDCs-based optoelectronics.

In the theoretical prediction, density functional theory (DFT) calculation of joint density of states (JDOS) was utilized to interpret the broad C peak in $MoS_2$ resulting from band nesting.[12-14] However, this simplified calculation cannot reproduce the sharp A, B exciton peaks because of the absence of the excitonic effect. Sole JDOS, a rough approximation of optical absorption, is not suitable to describe the multi-excitonic behaviors, especially for tungsten dichalcogenides (W-DCs) in Figure 1a. More advanced calculation like Bethe Salpeter equation (BSE) has been demonstrated by Qiu *et al* [15-17] and other groups [18-21] to yield more accurate absorption spectra. The calculated line shape of diverse excitons is highly consistent with experimental results. This indicates the necessity of BSE calculation to clarify the physical origin of various absorption peaks of TMDCs, a system dominated by strong electron-hole interactions.

Here, we employ momentum ($q$) resolved electron energy loss spectroscopy ($q$-EELS) to measure the dispersive behaviors of the intense higher-energy excitons at finite momenta.

Combined with the accurate BSE calculations, non-band-nesting transitions at Q valley and at midpoint of KM are found to be responsible for the high-energy peaks in W-DCs. Moreover, these high-energy transitions carry the specific spin polarizations, different from the well-known band nesting in molybdenum dichalcogenides (Mo-DCs). Our work will offer more insights to the physical origin of diverse excitons in 2D semiconductors and may inspire the manipulation of high-energy excitons with spin polarization in optoelectronic devices.

## RESULT AND DISCUSSION

*q-EELS of TMDCs.*

Monolayer TMDCs were prepared by chemical vapor deposition growth on silicon substrates, and then transferred onto Cu grids with lacey carbon film, and annealed at 300 ℃ in vacuum for 2 h to remove the surface residue before *q*-EEL spectra collection. The monochromated TEM (Triple C2 - JEOL ARM60MA) was operated at 60 kV, with an energy resolution of ~40 meV. The spectrometer entrance aperture (SEA) in diameter of 1 mm was used to select the specific *q* along the designated ΓK and ΓM directions, with a momentum resolution of 0.025 Å$^{-1}$. As illustrated in Figure 1a, the EEL spectra collected in the optical limit $q \rightarrow 0$ present sharp A, B excitons, and the broad C excitons in Mo-DCs split into several peaks ($C_0$, $C_1$, $C_2$, *etc*.) in W-DCs. It is obvious that the multi-peak feature indicates their diverse *k*-pace origins which are still unknown and to be determined (Figure 1b).

Following our paradigm of *q*-EELS of WSe$_2$ monolayer,[22] we extend this methodology to other TMDCs and focus on the unexplored high-energy excitons which are much more intense than A excitons. Figure 2a shows a typical momentum-energy (*q-E*) diagram of monolayer WS$_2$. For all monolayer TMDCs system, the experimental results show that sharp A excitons survive only for $q < 0.18$ Å$^{-1}$ ~ |ΓK|/8, which is a very small momentum range. The *q*-E diagrams in ΓK and ΓM directions show almost the same behaviors (Supplementary Figure S1) in the limited *q* range. This means there is no obvious in-plane anisotropy within this limited *q* range where excitons survive. Compared with the valence-splitting A, B excitons, the broad C peak with even higher intensity presents a much dispersive behavior.

Figure. 2b and Figure 2c show the detailed $q$ dependent EEL spectra of $WS_2$ and $WSe_2$ along ΓK direction (Figure S2 for $MoSe_2$). Note that the split C excitons ($C_1$, $C_2$, *etc*.) in W-DCs at $q \to 0$ turn into one broad peak as $q$ increases. To follow the naming tradition of excitons in Mo-DCs, we still use the same notation C in the description of exciton dispersions of W-DCs. This broad high-energy exciton presents obvious blueshift trend as $q$ increases, forming a more dispersive behavior than A exciton.

It is worthwhile to mention that the multi-C peaks in W-DCs were marked as A′, B′ excitons in Refs. 13, 22, 23 and assigned as the 2s Rydberg states of A or B excitons. In this assignment, the energy gap (Δ) of A to A′ excitons will be Δ=0.81 eV and 0.98 eV for $WSe_2$ and $WS_2$, respectively. In the 2D hydrogenic model with $E_n = -E_0/(n-0.5)^2$, this gap should be $\Delta = 4E_0 - 4E_0/9 = 32E_0/9 = 8E_b/9$, where $E_b=4E_0$ is exciton binding energy. Then one can infer $E_b$ ($WSe_2$) = 0.81/(8/9) = 0.91 eV and $E_b$ ($WS_2$) = 0.98/(8/9) = 1.10 eV. However, the values are much larger than the reported ~ 0.6 eV in GW-BSE calculation[16, 24] and scanning tunneling spectroscopy (STS) measurement.[17] Therefore, this discrepancy indicates the assignment of high-energy peaks as A′, B′ is not accurate and the multi-C peak in W-DCs should result from some other unknown transitions rather than the 2s Rydberg excitons.

*Dispersions of excitons.*

To quantify the dispersion behaviors of A, B excitons and post-gap C excitons, we summarize all the experimental dispersion data in Figure 3. All A, B excitons present parabolic dispersions as shown in Figure 3a. One can also find the splitting of A, B excitons in energy is 0.4 eV for both $WS_2$ and $WSe_2$, and decreases to 0.24 eV for $MoSe_2$ and to 0.18 eV for $MoS_2$. Based on the parabolic dispersions in Figure 3a, effective mass $m^*$ of A excitons can be inferred, as shown in Table 1. Exciton $m^*$ in Mo-DCs reaches 0.8 ~ 0.9 $m_e$ and decrease to 0.6 $m_e$ in W-DCs. This trend is also consistent with the GW calculated results[18] in Table 1, which indicates the difference of W-DCs with the stronger spin orbit coupling compared to the well-known Mo-DCs.

Compared to the A excitons, the intense C excitons present more dispersive relations with complicated non-parabolic trends (Figure 3b). Among all TMDCs, the broad C exciton of WSe$_2$ presents a small slope in the dispersion curves (in Figure 3b). This enables us to identify their possible distinctive *k*-space origins, interpreted by a further theoretical analysis later on.

*BSE loss functions.*

In theory, dispersions of excitons can be obtained by solving *q* dependent Bethe-Salpeter equation (BSE)[25-27] within Tamm-Dancoff approximation (TDA):

$$(E_{c,k+q} - E_{v,k})A^{\lambda,q}_{v,c,k} + \sum_{v'c'k'}\langle v,c,k,q|K^{eh}|v',c',k',q\rangle A^{\lambda,q}_{v',c',k'} = \Omega^{\lambda,q}A^{\lambda,q}_{v,c,k} \quad (1)$$

Here, $A^{\lambda,q}_{v,c,k}$ is the amplitude of the wavefunction of excitons comprised of electron at $|c,k+q\rangle$ state and hole at $|v,k\rangle$ state, $\Omega^{\lambda,q}$ is the energy of the $\lambda^{\text{th}}$ exciton, $K^{eh}$ is the electron-hole interaction kernel which includes both repulsive exchange term and direct term between electron-hole pairs. After solving BSE, we obtain optical absorption Im($\varepsilon$) and EELS loss function Im(-1/$\varepsilon$) at finite momentum *q*. It is worth mentioning that the optical form Im($\varepsilon$) and EELS form Im(-1/$\varepsilon$) are still different even at $q \to 0$, although the whole line shape will be quite similar. The EELS form usually present peaks blueshifted with respect to the optical form (Figure S3), also exemplified by the Kramers-Heisenberg dielectric function (in refs. 28, 29) $\varepsilon = \varepsilon_1 + i\varepsilon_2 = \varepsilon_\infty + \sum_j[\frac{\omega_{jp}^2(\omega_{jT}^2-\omega^2)}{(\omega_{jT}^2-\omega^2)^2+\omega^2\gamma_j^2} + i\frac{\gamma_j\omega_{jp}^2\omega}{(\omega_{jT}^2-\omega^2)^2+\omega^2\gamma_j^2}]$. Here $\omega_{jp}$, $\omega_{jT}$, $\gamma_j$ are the oscillator strength, transverse frequency (peak position), and damping constant of the *j*$^{\text{th}}$ excitation peak of the whole spectrum, respectively.

Figure 4 shows the experimental *q*-EEL spectra and BSE calculated loss function Im(-1/$\varepsilon$) of MoS$_2$. All the main features of BSE are well consistent with the experimental peaks in Figure 4a, except the different broadenings of C peak. Furthermore, the blueshift trend of all excitons as *q* increases agrees quite well with the experiment. Through tracking the q dependent BSE peak positions, one can obtain the theoretical dispersions of excitons. As shown in Figure S4, A, B excitons present standard parabolic dispersions. And high-energy C excitons are dispersed almost linearly, although the slope is lower than the experimental result in Figure 3. A more perfect quantitative matching needs much expensive computation costs and treatments such as lifetime broadening in the calculation, which is far beyond the scope of present work. The well matching of BSE and experimental spectra illustrates the accuracy of the exciton prediction and also the necessity of BSE in the classification of excitons, especially for the unknown high-energy excitons.

It should be mentioned that in both the experiments and BSE calculations one can never reach exactly $q = 0$ ($\Gamma$ point). We always use a finite and small $q$ to approach the optical limit $q \to 0$. Before reaching the threshold value of $q$, we still need to take the $q$ dependent EELS from $q \sim 0.1$ Å$^{-1}$ to $q < 0.02$ Å$^{-1}$. Then we know the EEL spectra at $q < 0.02$ Å$^{-1}$ can mimic the optical limit. Both $q$-EELS and BSE calculation at series of small $q$ are necessary to corroborate the peaks observed and explain the physical origins.

*Physical origin of high-energy absorptions.*

Excitons at $q \to 0$ is the dominating physical process in the optoelectronics of TMDCs, which is of both scientific and technological significance to the applications. To clarify the initial and final states involved, one needs to turn to the computational costly BSE calculation. Figure 5a and Figure 5b present the calculated BSE loss functions Im(-1/ε) of MoS$_2$ and WSe$_2$ at $q \to 0$ with spin-orbit coupling (SOC) included in the Kohn-Sham ground states. This will provide better accuracy than the perturbative treatment.[15, 30] All the main features from A exciton to (multi-) C exciton agree well with their experimental spectra.

As mentioned earlier, the multi-C peak feature in WSe$_2$ in the optical limit $q \to 0$ indicates the possible diversity of the transitions in the band structure. To track their $k$-space origins, one needs to calculate BSE and obtain eigenvectors $|A_\lambda\rangle = \sum_{v,c,k} A_{v,c,k}^{\lambda,q\to 0}|v,c,k\rangle$. Then the exciton wavefunction can be written as [31] $\Psi_k^{\lambda,q\to 0}(r_h, r_e) = \lim_{q\to 0}\sum_{v,c} A_{v,c,k}^{\lambda,q}\varphi_{v,k}(r_h)\varphi_{c,k+q}^*(r_e)$ where $\varphi_{v,k}(r_h)$ and $\varphi_{c,k}(r_e)$ are wavefunctions of hole and electron, respectively. From this exciton wavefunction, we derive a quantity "excitonic weight $\eta$", which defines the contribution of valence and conduction bands to the specific $\lambda^{th}$ exciton at any given $k$ point:

$$\eta_{v,k}^{\lambda} = \sum_c |A_{v,c,k}^{\lambda,q\to 0}|^2, \quad \eta_{c,k}^{\lambda} = \sum_v |A_{v,c,k}^{\lambda,q\to 0}|^2 \qquad (2)$$

Figure 5c shows the BSE C-excitonic weights $\eta$ of each band of MoS$_2$ and WSe$_2$, illustrated by the colored error bars of each band. Besides the 2D $k$-space wavefunction distribution (inset of Figure 5a), the $\eta$-weighted band structures in Figure 5c provide intuitive information such as the accurate initial and final states involved and also their spin polarizations. In MoS$_2$, the C-excitonic weight (light blue in Figure 5c) in the band nesting regions between $\Gamma$ and Q is consistent with the previous theoretical calculations.[12, 15, 21] While in WSe$_2$, the BSE shows the third peak C$_0$ in Figure 1a is actually B′ 2s-Rydberg state exciton and the second peak is actually the mixture of A′ and B excitons (Figure 5c). In higher-energy range, the C$_1$ peak of WSe$_2$ has the largest excitonic weight almost at Q valley (pink) and C$_2$ peak does between K and M (pink) in Figure 5c. Furthermore, one can see that C$_1$ peak is from the electronic transition of the first valence/conduction band at Q point, and that C$_2$ peak arises from the transition at midpoint of KM between the first valence band and the second conduction band. Notably, the bands responsible for C$_2$ peak are quite unparallel (right panel of Figure 5c), in contrast to the known band-nesting mechanism in MoS$_2$. In analogy with WSe$_2$, the BSE calculated spectrum and excitonic weights of WS$_2$ (Figure S5) also show very similar $k$-space origins for the multi-C high-energy peaks. The specific transitions at Q and midpoint of KM (Figure S5) contributes to the intense high-energy absorptions of WS$_2$.

A further check of the spin-resolved electronic structures shows that the both bands involved in C$_2$ peak of WSe$_2$ are spin-down polarized (in purple in Figure 5c), in analogy with the spin polarization of A exciton.[3] This spin-down polarization also persists in the C$_1$ peak at Q valley. When it turns to C exciton in MoS$_2$, both bands in the band-nesting region are almost "spin-degenerate" (left panel of Figure 5c). Intuitively, this spin polarization of C peaks in WSe$_2$ should arise from the intrinsic strong SOC effect in W-DCs. This also indicates the possible valleytronics-like manipulation of intense high-energy excitons in W-DCs through the degree of freedom of spins.

To understand the contrasting multi-C absorptions of W-DCs with respect to Mo-DCs, one needs to check the mechanisms accounting for the excitonic weights. This difference results from the atomic orbitals. Group theory analysis shows that the highest occupied molecular orbital (HOMO) at $K_v$ (valence band maximum) consists of 0.49 $d_{xy}$ + 0.49 $d_{x^2-y^2}$ for both Mo-DCs and W-DCs,[32, 33] which belongs to E′ irreducible representation of $D_{3h}$ point group. While the lowest unoccupied molecular orbital (LUMO) at $K_c$ (conduction band minimum) changes from 0.94 $d_{z^2}$ ($A_1'$ irrep of $D_{3h}$) for Mo-DCs to 0.38 $d_{xy}$ + 0.36 $d_{x^2-y^2}$ + 0.23 $d_{z^2}$ for W-DCs.[33] This increase of the weight of $d_{xy}$ and $d_{x^2-y^2}$ (E′ irrep of $D_{3h}$) at $K_c$ of W-DCs will also persist into other points between K and M. For non-special point Q and KM midpoint of $C_s$ point group, the d-d transition of more $d_{xy} + d_{x^2-y^2}$ states (A′ irrep of $C_s$) is allowed in W-DCs. Compared to the case of Mo-DCs, the redistribution of $d_{xy} + d_{x^2-y^2}$ or $d_{z^2}$ in the first conduction band possibly increase the weight of the optical transition $C_1$ at Q point and $C_2$ between K and M in W-DCs. On the other hand, the larger SOC-induced band splitting in W-DCs may also give rise to the splitting of the high-energy optical transitions, which account for their multi-C features.

In the high energy end of $q$-EELS, π-like and π+σ-like plasmons appear at 8 ~ 25 eV and at a relatively large $q$. These plasmons are strongly suppressed at small $q$ and in energy window 0 ~ 4 eV which is of our interest. Therefore, the optical features we discussed above are not affected by the plasmons and result from the intrinsic band transitions. In the lower energy end, phonons of $MoS_2$ present an energy ≤ 50 meV, which is beyond the capability of our TEM. In the future, a 5-meV monochromator may make it feasible the $q$-EELS phonon dispersion measurement of TMDCs.

## CONCLUSIONS

In summary, we have presented a $q$-EELS technique to directly measure the dispersions of various excitons of monolayer TMDCs. For the most intense high-energy absorptions, we located their different *k*-space origins by a BSE-experiment joint analysis. In contrast with the band nesting mechanism in Mo-DCs, direct transitions at Q valley and at midpoint of MK with specific spin polarization should account for the high-energy absorptions in W-DCs. This implies that non-band-nesting mechanism play a more important role in the multi-C absorptions of W-DCs. Our work provides a fundamental perspective to visualize the exciton physics in 2D semiconductors, and will inspire more efforts to explore the whole spectrum covering all energies. Our *q*-EELS methodology can be generalized to any gapped or non-gapped crystal material to help to understand the spectral features. The discovery of most intense multi-C peaks at 3~3.5 eV, in the violet region of solar spectrum, will benefit us in harvesting high-energy excitons with spin polarizations in TMDCs' optoelectronic applications.

## METHODS

**Sample preparation.** High-quality monolayer TMDCs are prepared by chemical vapor deposition on the silicon substrate, and then transferred onto TEM grids with holey carbon film using spin-coated Poly (methyl methacrylate) (PMMA). Then the PMMA layer is removed using acetone, followed by isopropanol to dissolve possible organic residues. To further reduce the surface carbon or contamination, the samples are heated at 300 °C for 2 h before the TEM and EELS experiments.

***q*-EELS Experiment.** All *q*-EEL spectra are collected in standard diffraction mode in the TEM (Triple C2) operated at 60 kV, with an energy resolution of ~ 40 meV and momentum resolution of 0.025 Å$^{-1}$. Dual EELS is used to obtain the low loss signal and zero loss peak (ZLP) simultaneously with a dispersion of 0.005 or 0.01 eV/channel. Hence the ZLP drift can be easily calibrated and the stability or accuracy of the peak positions is $\leq \pm 0.01$ eV.

**Theoretical calculation.** All Density Functional Theory (DFT) calculations are performed by the Exciting package[26, 31] with energy cutoff of 60 Ry. We use 15 Å vacuum to avoid interactions between neighboring slabs. Atomic geometries of $MoS_2$, $MoSe_2$, $WS_2$ and $WSe_2$ are relaxed by optB86-VdW functional[34] with force and total energy convergence criterion of $10^{-5}$ and $10^{-7}$ Ry in atomic unit, respectively. Ground state electron density is calculated using PBE exchange correlation functional[35] with 50×50×1 *k*-sampling of Brillion zone. A scissor operator was used to mimic the GW quasiparticle energy correction like the other code GPAW and Yambo to reduce the computational cost. We cut dielectric matrix at 35 Ry on 18×18×1 uniform $q$-mesh in Brillion zone. 2D truncation of Coulomb interaction is employed to avoid interactions with the image slab. The Bethe-Salpeter equation (BSE) calculation uses the Kohn-Sham ground states with SOC considered, instead of the perturbative treatment.[15, 30] In the calculation of BSE matrix, 8 valence bands and 8 conduction bands are used. We consider only spin-singlet (S=0) BSE where both exchange and direct terms of the BSE Hamiltonian are fully taken into consideration. The spin-triplet (S=1, spin-flipping) BSE is not considered, as it is not dipole allowed and should be absent in the optical pumping[5] or EELS experiments at $q \rightarrow 0$. The loss function calculated from BSE exciton energies and oscillator strengths is convoluted by an artificial broadening of 50 meV. Finally, reciprocal-space excitonic weights are projected onto the band structure using BSE eigenvectors. The absorption spectra and $k$-space excitonic weights are also corroborated by another independent code Yambo.

## AUTHOR INFORMATION

### Corresponding Author

*E-mail: jinhuahong436@gmail.com (J. H.) or suenaga-kazu@sanken.osaka-u.ac.jp (K. S.)

### Author Contributions

J.H. and K.S. conceived this work. J. H., R. S. and K.S. performed the $q$-EELS measurement; M. K. and J. H. did the BSE theoretical calculation; H. X. contributed specimens; J.H., M. K., K.S., T.P. comment and analyze the data and theory; all authors cowrote the paper with full discussion.

### Notes

The authors declare no competing financial interest.

## ACKNOWLEDGMENTS

This work was supported by JST-CREST (JPMJCR20B1, JPMJCR20B5, JPMJCR1993) and JSPS-KAKENHI (JP16H06333, JP19K04434 and JP17H04797) and A3 Foresight program. H. Xu acknowledges the support from the National Natural Science Foundation of China (51972204). We thank Dr. Y. Li, Dr. S. Y. Quek, and Dr. F. Xuan for fruitful discussions and comments.

## ASSOCIATED CONTENT

### Supporting Information

Supporting Information is available free of charge on the ACS Publications website: Experimental *q-E* diagram of WS$_2$ and *q*-EELS of MoSe$_2$ (Fig. S1-S2); theoretically calculated optical absorption, loss function, exciton dispersion, excitonic weights, oscillator strength of TMDCs (Fig. S3-S6); removal of zero-loss peak background in EELS (Fig. S7).

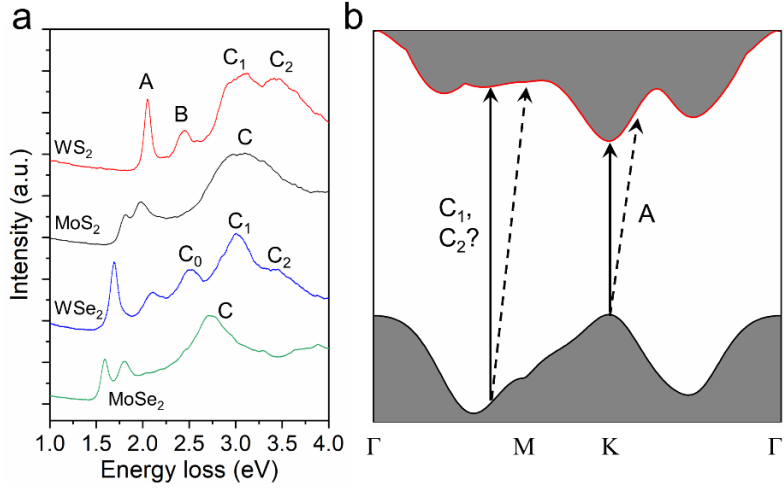

**Figure 1. Exciton fine structures of TMDCs and their schematic origins. (a)** Diverse excitons in monolayer TMDCs measured by $q$-EELS at $q \to 0$. For all TMDCs, the first two sharp peaks are A, B excitons resulting from valence band splitting. In the higher-energy range, the broad C peaks in Mo-DCs split into several peaks ($C_1$, $C_2$) in W-DCs. **(b)** Schematic transitions with zero-$q$ and nonzero-$q$ excitations. Here the origin of high-energy peaks is to be determined in $k$ space.

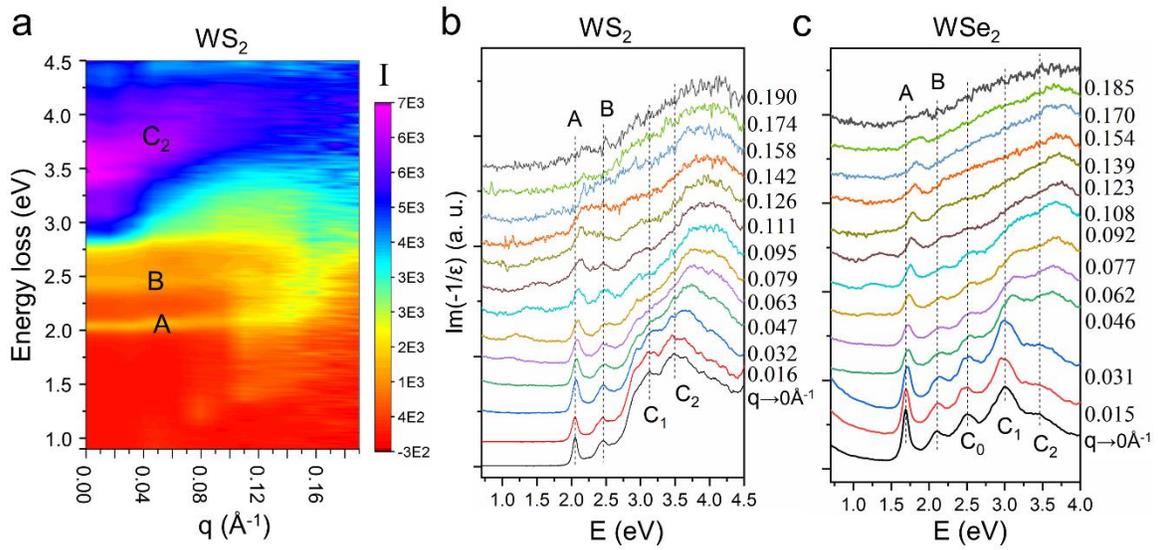

**Figure 2. Experimental *q-E* diagram and *q*-EELS of W-DCs. (a) Typical *q-E* diagram of monolayer $WS_2$ to show the dispersions of various excitons. (b, c) The detailed *q*-EEL spectra of $WS_2$ and $WSe_2$ along ΓK direction. Here the spectra of $WSe_2$ are from the superposition of two *q*-EELS series, which gives good signal-to-noise ratio.**

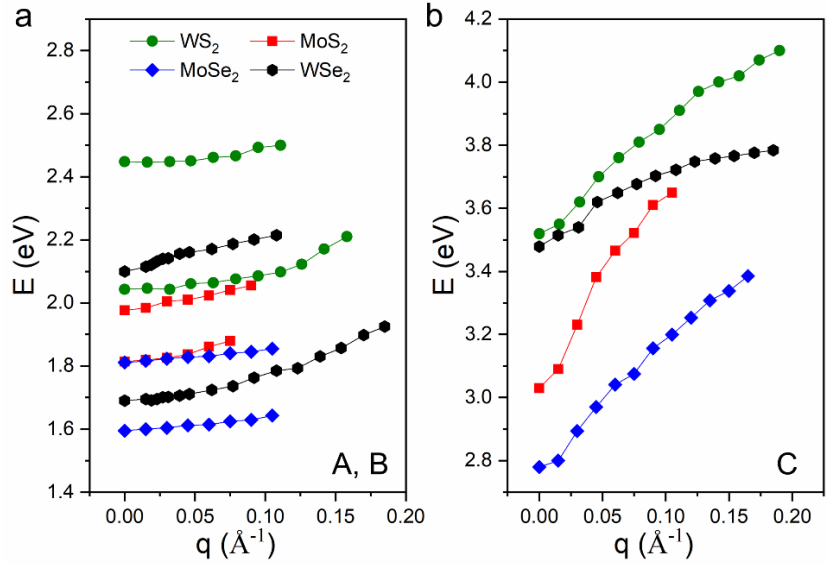

**Figure 3. Experimental dispersions of excitons. (a)** Dispersions of valley excitons (A, B). **(b)** Dispersions of post-gap peaks C of monolayer TMDCs. Since the split C excitons ($C_1$, $C_2$, *etc.*) in W-DCs at $q \to 0$ turns into one broad peak as $q$ increases, we still use the same notation C as Mo-DCs in the description of exciton dispersions. Compared to parabolic dispersed A, B excitons, the intense broad excitons present a much more dispersive but complicated behavior.

**Table 1 | Comparison of exciton effective mass ($m^*$) of monolayer TMDCs by $q$-EELS and GW-BSE.**

| $m^*/m_e$ | $MoS_2$ | $MoSe_2$ | $WS_2$ | $WSe_2$ |
| --- | --- | --- | --- | --- |
| $q$-EELS | 0.80±0.14 | 0.96±0.07 | 0.61±0.04 | 0.65±0.04 |
| GW-BSE[18] | 1.01 | 1.07 | 0.7 | 0.72 |

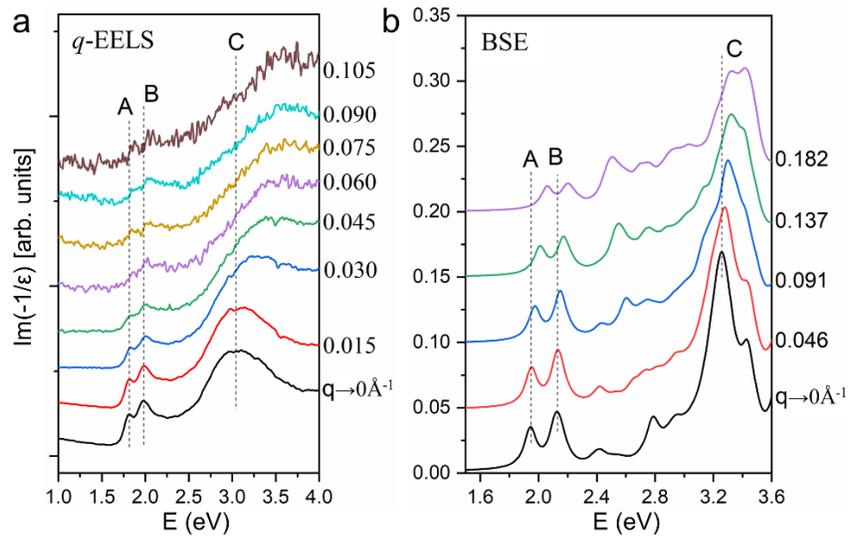

**Figure 4. Comparison of experimental and theoretical loss functions of MoS$_2$. (a)** *q*-EELS along ΓK direction. **(b)** BSE-calculated loss functions along ΓK direction. The BSE calculations well interpret the *q*-dependent blueshift of excitons.

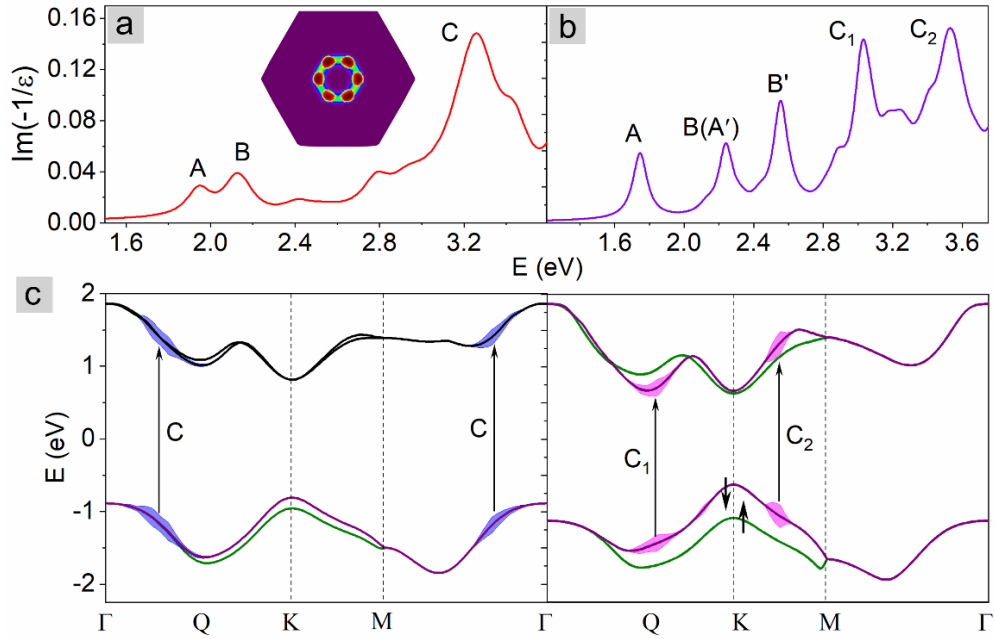

**Figure 5. Theoretical loss function Im(-1/ε) at $q\rightarrow 0$ and excitonic weights. (a, b) Loss functions of MoS$_2$ and WSe$_2$ with SOC included. The inset of a is the 2D projected $k$-space wavefunction distribution of C exciton. (c) BSE calculated $k$-point dependent excitonic weights projected onto each band responsible for the higher-energy excitons. The colored error bars of each band calculated with SOC in c) illustrates the excitonic weights. Bands in green are spin-up polarized and bands in purple are spin-down. All bands between Γ and M are spin-degenerate. Only the first two valence bands and two conduction bands are drawn in the band structures, since deeper bands contribute the excitonic weights at least two orders of magnitude lower.**

**TOC**

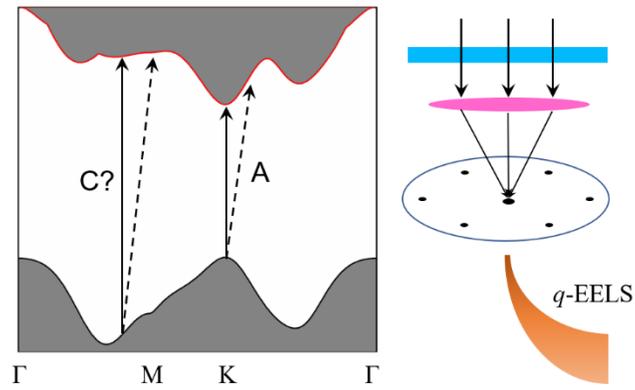

**The most intense high-energy absorption peaks of tungsten dichalcogenides were probed by momentum resolved EELS and proved to result from multi-transitions at non-band-nesting points.**